\documentclass{elsart}

\usepackage{epsfig}
\usepackage{natbib}
\usepackage{amssymb}
\usepackage{latexsym} 
\usepackage{mathptm}

\def\beq{\begin{equation}}
\def\eeq{\end{equation}}
\def\beqn{\begin{eqnarray}}
\def\eeqn{\end{eqnarray}}

\begin{document}

\begin{frontmatter}

\title{Anti-Gravitation}

\author{S.~Hossenfelder}
\ead{sabine@physics.ucsb.edu}
\address{Department of Physics, University of California\\Santa Barbara, CA 93106-9530, USA}

\begin{abstract}
The possibility of a symmetry between gravitating and 
anti-gravitating particles is examined. The properties of the anti-gravitating fields 
are defined by their behavior under general diffeomorphisms. The equations of 
motion and the conserved canonical currents are derived, and it is shown that the kinetic 
energy remains positive whereas the new fields can make a negative contribution
to the source term of Einstein's field equations. The interaction between the
two types of fields is naturally suppressed by the Planck scale.  
\end{abstract}

\begin{keyword}
Negative Energy, Lorentz-Transformations
\PACS 04.20.Cv \sep 11.30.-j \sep 11.30.Cp 
\end{keyword}

\end{frontmatter}

\section{Introduction}
  
The role of symmetries in nature has marked the progress of physics during the last centuries. 
The use of symmetries has proven to be a powerful tool like no other, underlying
General Relativity (GR), leading to the discovery of anti-particles and the establishment of the
quark model. 

With these successes of symmetry principles in mind, it is a question lying at hand 
whether there is a symmetry between positive and
negative masses.

In this work,  a framework is presented to include particles with negative gravitational
charge into classical gravity and field theory.  The anti-gravitating particles are 
introduced as particles that transform in a modified way under general diffeomorphisms. In a
(locally) flat space, they transform like the standard particles. This  generalization
of the equivalence principle leads to the introduction of a modified covariant derivative.
The generalized covariant derivative ensures the homogeneous
transformation behavior of the new field's derivatives. From this, the equation of motion is derived
for the new fields.

The introduction of anti-gravitation in this way solves three severe problems that seem to come 
along with the notion of negative masses: 

\begin{enumerate}
\item
In GR, the motion of a particle in a gravitational 
field is independent of the particle's mass. This seems to indicate that no particle
can be repelled by a positvely gravitating source and leads to immediate contradictions.
In the here discussed setup, also the anti-gravitating particle's
equation of motion is modified, which follows from its
behavior under general diffeomorphisms. In this case, like gravitational charges attract, 
and unlike charges repel.

\item 
In the here used formalism, only the source to the gravitational
field can experience a change of sign. The kinetic energy 
term in the Lagrangian remains positive and thus, the vacuum remains stable.

\item 
The third problem is the lacking evidence. As we will show, the interaction between gravitating and anti-gravitating
matter is mediated by gravity only, and therefore is suppressed by the large value of the Planck scale.
\end{enumerate}

In addition to being a viable extension of {\sc GR}, the here proposed existence of negative
gravitational charges can open new points of view in Cosmology and Astrophysics, especially
with regard to singularity avoidence and Dark Energy.  Though it will become clear from the
analysis that the
newly introcued particles are only very weakly interacting with standard matter, their presence
might be relevant at large (cosmological) distances, at high (Planckian) densities and in
strongly curved backgrounds.

The proposal of a gravitational
charge symmetry has previously been examined in Refs. \cite{linde1,linde2,Moffat:2005ip,Kaplan:2005rr}. 
Furthermore, there have been various approaches \cite{Bondi,Quiros:2004ge,Borde:2001fk,Davies:2002bg,Ray:2002ts,
Rosenberg:2000cv,Torres:1998cu,Zhuravlev:2004vd,Faraoni:2004is} 
to include anti-gravitating matter into quantum 
theory as well as into GR. The topic of negative energies has recently received attention within the context
of  ghost condensates \cite{Arkani-Hamed:2003uz,Arkani-Hamed:2003uy,Arkani-Hamed:2005gu,Mann:2005jz,Krause:2004bu}.  
  
This paper is organized as follows: In the next section, the definition for the anti-gravitating
particles is given, based on their transformation properties under the gauge- and Lorentz-group. 
Section three examines the modified field equations that follow from the inclusion of these
particles into {\sc GR}, as well as conservation laws, and the motion of test particles. General
comments and discussion can be found in section \ref{discuss}. We conclude in section \ref{concl}.

Throughout this paper we use the convention $c=\hbar=1$. The signature of the metric is $(-1,1,1,1)$. Small
Greek indices $\kappa,\nu,\epsilon...$ are space-time indices.
  
\section{Definition of Anti-Gravitating Matter}

The unitary representations $U$  of a gauge group $G$ define the transformation behavior of particle fields $\Psi$. For
two elements of the group $g,g'$, the representation fulfils
\beqn
U(g) U(g') = U(gg') \label{multrep}\quad,
\eeqn
and the field transforms as $\Psi \to \Psi' = U(g) \Psi$.
From this, it is always possible to construct a second representation, defined by 
\beqn
\widetilde{U}(g) = U((g^T)^{-1}) \quad,
\eeqn
which belongs to the charge-conjugated particle. The anti-particle $\overline{\Psi}$
transforms according to the contragredient representation, ${\overline{U}}$, which is
${\overline U}(g) = U(g^{-1})$.   

 In case of a local symmetry, these transformations lead to the introduction of
gauge-covariant derivatives in the usual way. Suitable combinations of particles with anti-particles allow
to construct gauge-invariant Lagrangians.

From the above, it is clear why there is no charge-conjugation for gravity. If the gauge-group is the 
Lorentz-group $SO(3,1)$, then the elements $\Lambda$ fulfill
$\Lambda^{-1} = \Lambda^T$,
which means that in this case the second representation $\widetilde{U}$ is equivalent to $U$. 

However, this does not apply when the field transforms under a general diffeomorphism $G$. Let $\Psi$ be
a vector field and an element of the tangent space $TM$. Under a general diffeomorphism $G$, the fields and
its conjugate behaves as
\beqn
TM~ &:& \Psi \to \Psi' =  G \Psi \quad,\quad 
{TM^*} :  {\overline \Psi} \to {\overline \Psi'} = {\overline \Psi} G^{-1} \quad, \label{stdtrafo}
\eeqn
where $TM^*$ is the dual to $TM$. 

The equivalence principle requires that the fields on our manifold
locally transform like in Special Relativity. I.e. if $G$ is an element of the Lorentz-group the fields have
to transform like Lorentz-vectors. However, the generalization to a general diffeomorphism is
not unique. Instead of Eq. (\ref{stdtrafo}) one could have chosen the field to transform according to
\beqn
{\underline{TM}}~ : {\underline \Psi} \to {\underline \Psi}' = (G^T)^{-1} \underline\Psi  \quad,\quad
{\underline{TM}}^* &:&   {\underline {\overline \Psi}} \to {\underline {\overline \Psi'}} = 
{\underline {\overline \Psi}} G^T \quad. \label{trafo}
\eeqn
Here, the space ${\underline{TM}}$ is a vector-space which spans the basis for these fields, and ${\underline{TM}}^*$ is
its dual. In case $G$ was an element of the Lorentz-group, i.e. $G^{-1}=G^T$, both representations 
(\ref{stdtrafo},\ref{trafo})
agree. For general diffeomorphisms that will not be the case. Indeed, one sees that the newly introduced fields 
will have a modified scaling behavior. 
 
For the following, it is convenient to introduce a map $\tau$ which, in the vector-representation, is a 
vector-space isomorphism from $TM$ to ${\underline{TM}}$. For the map $\tau: \underline \Psi = \tau \Psi$ 
to transform adequately $\underline \Psi' = \tau' \Psi'$ one finds the behavior
\beqn
\tau' = (G^T)^{-1} \tau G^{-1} \label{trafotau} \quad.
\eeqn

\begin{figure}[t]
\vspace*{-0.0cm}
\centering \epsfig{figure=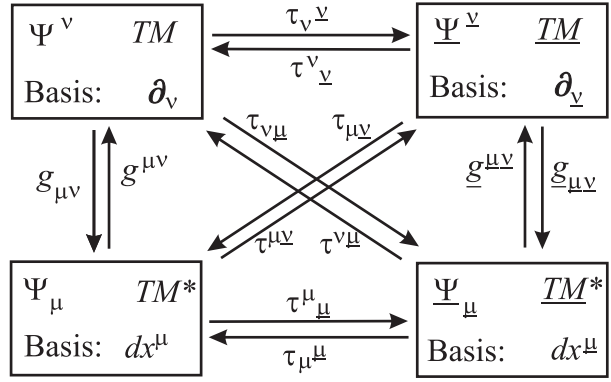, width=8cm}
 
\caption{Relations between the maps. The left side depicts 
the  tangential and co-tangential spaces. The right side depicts the corresponding
spaces for the anti-gravitational fields.\label{fig1}}
\end{figure}

It will be useful to clarify the emerging picture of space-time properties by having a close
look at a contravariant vector-field $\Psi^{\kappa}$ as depicted in Figure \ref{fig1}. This field 
is a cut in the tangent bundle, that is the set of tangent spaces $TM$ at every point of 
the manifold which describes our space-time.
The field is mapped to its covariant field, $\Psi_{\nu}$, a cut in the co-tangent bundle, $TM^*$, by the 
metric tensor $\Psi_{\nu} = g_{\kappa \nu} \Psi^{\nu}$. 


The newly introduced field  
${\underline \Psi}^{\underline \kappa}$ (from here on named anti-gravitating)
transforms under local Lorentz-transformations like a Lorentz-vector in Special Relativity. But it 
differs in its behavior under general diffeomorphisms. 

In order not to spoil the advantages of the Ricci-calculus, it will be useful to introduce a basis
for the new fields that transforms accordingly.  Locally, the new field can be expanded in this 
basis ${\partial}_{\underline \kappa}$. These basis elements form
again a bundle on the manifold, that is denoted with ${\underline{TM}}$. To each of the elements of
 ${\underline{TM}}$ also a dual space exists, defined as the space of all linear maps on ${\underline{TM}}$. This
 space is denoted by ${\underline{TM}}^*$ and its basis as $d{{x}^{\underline \kappa}}$.
The map from ${\underline{TM}}$ to ${\underline{TM}}^*$ will be denoted 
${\underline g}_{{\underline \kappa}{\underline \nu}}$, and defines a scalar product on ${\underline{TM}}$. 

Note, that the underlined indices on these quantities do not
refer to the coordinates of the manifold but to the local basis in the tangential spaces. All of these
fields still are functions of the space-time coordinates $x_\nu$.

$\tau$ maps the basis of one space into the other. We can expand it as 
$\tau = \tau_{\nu}^{\;\;{\underline \kappa}} ~ d x^\nu \partial_{\underline \kappa}$, 
or $\tau = \tau^{\nu}_{\;\; {\underline \kappa}} ~ d x^{\underline \kappa} \partial_\nu
$ respectively, where we have introduced the inverse functions by
\beqn
\tau_{\nu}^{\;\; {\underline \kappa}} \tau^{\nu}_{\;\;{\underline \epsilon}} 
= \delta^{\underline \kappa}_{\;\;\underline \epsilon} \quad,\quad 
  \tau^{\epsilon}_{\;\; {\underline \kappa}} \tau_{\nu}^{\;\; {\underline \kappa}} = \delta^{\epsilon}_{\;\;\nu}\quad. 
\eeqn
For completeness, let us also define the 
combined maps:
\beqn
TM \longrightarrow {\underline{TM}}^* &:& \quad \tau_{\nu \underline \nu} = \tau_{\nu}^{\;\; {\underline \kappa}} 
{\underline g}_{{\underline \kappa}{\underline \nu}} \quad,\quad
{{TM}}^* \longrightarrow {\underline{TM}} : \quad \tau^{\nu \underline \nu} =  
g^{\nu \kappa} \tau_{\kappa}^{\;\; \underline \nu}   \quad.
\eeqn
As one sees from the transformation properties summarized in
Eqs.(\ref{trafo}), the map $\tau_{\nu \underline\nu}$ maps an element of $TM$
to its transponed. This transposition does not involve the metric tensor and is independent of
the metric itself. In particular, one sees immediately that the determinant $|\tau_{\nu \underline \nu}| =1$, since
transposition does not change the determinant. 

Basis transformations in both spaces belong together, since
their elements share the transformation properties Eqs.(\ref{trafo}) with the same $G$.  
The map $\tau_{\nu \underline \nu}$ does not introduce an additional
coordinate transformation in the underlined space, it does not change the volume element, introduce a shear, or
an additional rotation on ${\underline{TM}}^*$. One should keep
in mind that the introduction of the additional spaces is just a helpful tool to deal with the different
transformation properties of the new fields.

The relation between the introduced quantities is summarized in Figure \ref{fig1} from which we also 
read off the cycle leading back to identity
on $TM^*$ and on ${\underline{TM}}^*$, respectively
\beqn
\tau^{\kappa {\underline \nu}} {\underline g}{}_{{\underline \kappa} {\underline \nu}}  
\tau^{\nu {\underline \kappa}} g_{\nu \epsilon}    = \delta^\kappa_{\; \epsilon} \quad, \label{cycle} \quad
{\underline g}{}^{{\underline \kappa} {\underline \nu}} \tau_{\nu {\underline \kappa}} g^{\kappa \nu} \tau_{\kappa {\underline \epsilon}} = \delta^{\underline \kappa}_{\; {\underline \epsilon}} \quad.
\eeqn
It is further helpful to note that $\tau$ can be expressed in a more intuitive form. Writing 
$\partial/\partial x^\nu = (\partial x^{\underline \kappa} / \partial x^\nu )~ \partial / \partial x^{\underline \kappa}$ one
identifies
\beqn
\tau^{\nu}_{\;\;\underline\kappa} = \frac{\partial x^\nu}{\partial x^{\underline \kappa}} \quad, \label{tau}
\eeqn
where we should keep in mind that $\partial_{\underline \kappa}$ is a just the basis in ${\underline{TM}}$ and
does not correspond to an actual 'direction' on the manifold as $\partial_{\nu}$ does.

The coordinate expansion of $\tau$ seems to depend on the basis chosen in ${TM}$ as well as on those
in  ${\underline{TM}}$. However,
this ambiguity in the form of the map is only seemingly. The basis $\partial_\nu$ and ${\partial}_{\underline \nu}$ 
does not transform independently as stressed earlier. 
The relation between both basis transformations is fixed 
by the map $\tau$. To see this, note that even though the map $\tau$ for certain choices of coordinates
might be trivial, its importance lies in its transformation behavior. 
Its covariant derivative takes into account the change of transformation behavior from $TM$ to ${\underline{TM}}$. 

The transformation behavior in Eq. (\ref{trafotau}) together with the observation 
that in a locally orthonormal basis both fields transform identical under 
local Lorentz-trans\-formations, 
gives us an explicit way to construct $\tau$. We choose a local orthonormal basis $\hat e$ in $TM$, 
which is related to the coordinate basis by the locally linear map $E \hat e = \partial$. In this basis, 
the metric is just $\eta$ and $\hat \tau$ is just the identity. One then finds $\tau$ in a general coordinate
system by applying Eq.(\ref{trafotau})
\beqn
\tau = (E^T)^{-1} \hat \tau E^{-1} = (E E^T)^{-1}  \quad \label{tauE}.
\eeqn
Using Eq. (\ref{tauE}), we again confirm that the determinant 
is $|\tau_{\kappa {\underline \kappa}}|=1$ and therefore 
\beqn
1 &=& g {\underline g} \quad,\quad |\tau_\kappa^{\;\; \underline \kappa}| = |g| 
\quad,\quad |\tau^\kappa_{\;\; \underline \kappa}| = |\underline g| \quad. \label{gdg} 
\eeqn

The properties of the vector-fields are transferred directly to those of fermionic fields by using the fermionic
representation and transformations of $\tau$. In this case, 
the map $(\;)^\dag \gamma^0$, instead of the metric, is used to relate a particle to the particle transforming under 
the contragredient representation.


It is now straightforward to introduce a covariant derivative for the new fields, in much the same way 
as one usually introduces the derivative for the charge-conjugated particles. 
We will use the notation $\nabla$ for the general-covariant derivative and
$D$ for the general covariant derivative including the gauge-derivative of the fields. It is understood
that the form of the derivative is defined by the field it acts on, even though this will not
be noted explicitly. 

Let us first introduce the derivative in the direction of $\nu$ on the basis in a 
general way by
\beqn
\nabla_\nu \partial_{\kappa} &=& \Gamma^{\epsilon}_{\; \nu \kappa} ~ \partial_\epsilon  \quad,\quad
{\nabla}_{\nu} {\partial}_{\underline \kappa} = 
\Gamma^{{\underline \epsilon}}_{\;\; \nu {\underline \kappa}}~ 
{\partial}_{\underline \epsilon} \quad. \label{stdder}
\eeqn
We will further also use the notation 
\beqn
\nabla_{\underline \kappa} = \tau^\kappa_{\;\;\underline \kappa}\nabla_\kappa \quad \label{DnuD}\quad.
\eeqn
From Eqs. (\ref{stdder}) it follows using the Leibniz-rule that
\beqn
\nabla_\nu dx ^ {\kappa} &=& - \Gamma^{\kappa}_{\; \nu \epsilon} dx^\epsilon \quad,\quad
{\nabla}_{\nu} dx ^{{\underline \kappa}} =  
- \Gamma^{{\underline \kappa}}_{\;\; \nu {\underline \epsilon}}~ 
dx ^{\underline \epsilon} \label{stdder4}
\quad.
\eeqn
So far, this is only a definition for the symbols $\Gamma^{\epsilon}_{\; \nu \kappa}$ 
and $\Gamma^{{\underline \epsilon}}_{\;\; \nu {\underline \kappa}}$. 
To get an explicit formula for these Christoffelsymbols one commonly uses the requirement that the
covariant derivative on the metric itself vanishes. The geometrical meaning is that the scalar product
is covariantly conserved. To keep the symmetry between both spaces, we require that also the scalar
product of the new fields is conserved, that is
$
\nabla_\lambda g_{\nu \kappa} =0$ and $\nabla_{\lambda} {\underline g}_{\underline \nu \underline \kappa} = 0$.
Or, since $|\tau^{\lambda}_{\;\;\underline\lambda}| \neq 0$, the latter is with Eq.(\ref{DnuD}) equivalent to 
$\nabla_{\underline \lambda} g_{\underline \nu \underline \kappa}=0$. 
From this one finds in the usual way (see e.g. Ref.\cite{Weinberg}) 
\beqn
\Gamma^{\nu}_{\;\; \lambda \kappa} &=& \frac{1}{2} g^{\nu \alpha} 
\left( \partial_\lambda g_{\kappa\alpha} + \partial_\kappa g_{\lambda \alpha} 
- \partial_\alpha g_{\kappa \lambda} \right) \quad, \label{chrisnorm} \\
\Gamma^{\underline\nu}_{\;\; \underline\lambda \underline\kappa} &=& \frac{1}{2} {\underline g}^{\underline\nu \underline \alpha} 
\left( \partial_{\underline \lambda} {\underline g}_{\underline{\kappa\alpha}} +  
\partial_{{\underline\kappa}} {\underline g}_{\underline{\lambda \alpha}} 
- \partial_{\underline\alpha} {\underline g}_{\underline\kappa \underline\lambda} \right) \quad.
\eeqn
Using that the Christoffelsymbols transform homogenously in the second index and removing the underlined derivatives
we find
\beqn
\Gamma^{\underline\nu}_{\;\; \lambda \underline\kappa} &=& \frac{1}{2} {\underline g}^{\underline\nu \underline \alpha} 
\left( \partial_{\lambda} {\underline g}_{\underline{\kappa\alpha}} +  
\tau_{\lambda}^{\;\;\underline \lambda} \tau^{\kappa}_{\;\;\underline \kappa} \partial_{{\kappa}} 
{\underline g}_{\underline{\lambda \alpha}} 
- 
\tau_{\lambda}^{\;\;\underline \lambda} \tau^{\alpha}_{\;\;\underline \alpha}
\partial_{\alpha} {\underline g}_{\underline\kappa \underline\lambda} \right) \label{chrisag} \quad.
\eeqn
From this, one can explicitly compute the form of the derivatives in Eqs. (\ref{stdder}). 

At this point it is useful to state a general expectation   about the connection coefficients. 
For a particle moving in a curved spacetime, it is possible to choose a freely falling coordinate system, in which
the Christoffelsymbols in Eq.(\ref{chrisnorm}) vanish. However, this freely falling frame for the particle will in general
not also be a freely falling frame for the anti-gravitational particle. Therefore, the Christoffelsymbols in Eq.(\ref{chrisag}) will
not vanish in the freely falling frame of the usually gravitating particle. Both sets of symbols therefore will
not be proportional to each other.

One further derives the useful relations
\beqn
\nabla_\nu \tau_{\alpha \underline \alpha} &=& 
\partial_\nu \tau_{\alpha \underline \alpha} - 
\Gamma^{\kappa}_{\;\; \nu \alpha } \tau_{\kappa \underline \alpha} - 
\Gamma^{\underline \kappa}_{\;\; \nu \underline \alpha} \tau_{\alpha \underline \kappa} \\
\nabla_\nu \tau^{\alpha}_{\;\; \underline \alpha} &=& 
\partial_\nu \tau^{\alpha}_{\;\; \underline \alpha}  + 
\Gamma^{\alpha}_{\;\; \nu \kappa} \tau^{\kappa}_{\;\;\underline \alpha} 
- \Gamma^{\underline \kappa}_{\;\; \nu \underline \alpha} \tau_{\alpha \underline \kappa} \label{dtau} \quad.	
\eeqn
With these covariant derivatives, one obtains the Lagrangian of the anti-gravitational 
field by replacing all quantities with the corresponding anti-gravitational quantities
and using the appropriate derivative for the new fields to assure homogenous transformation behavior
\beqn
{\underline {\mathcal L}}  = 
{{\mathcal L}}(g_{\kappa \nu} \to {\underline g}{}_{{\underline \kappa} {\underline \nu}} ,  
\Psi \to {\underline \Psi}) \quad, \quad
{{\mathcal L}}  = 
{\underline {\mathcal L}}({\underline g}{}_{ {\underline\kappa} {\underline{\nu}}} \to g_{\kappa \nu} ,  
{\underline \Psi} \to {\Psi}) \quad. \label{llu}
\eeqn
 
To start with the most important example, the Lagrangian of fermionic fields can now be composed from the
new ingredients as ${\mathcal L}^{\rm tot}_{F} = {\mathcal L}_F + {\underline {\mathcal L}}_F$ with 
\beqn
{\mathcal L}_F &=& ( { D}\hspace*{-.25cm}/\hspace*{.15cm} \overline{\Psi} )  \Psi + 
{\overline \Psi} ( D\hspace*{-.25cm}/\hspace*{.15cm} \Psi )   \quad,\quad
{\underline {\mathcal L}}_F = 
( { {D}}\hspace*{-.25cm}/\hspace*{.15cm} {\underline{\overline{\Psi}}} ) {\underline \Psi}  
+ { {\overline{ \underline \Psi}}} ( { {D}}\hspace*{-.25cm}/\hspace*{.15cm} {\underline{{\Psi}}} )  \quad.
\eeqn
All other mixtures do not
obey general covariance and/or gauge covariance. 

The Lagrangian for  anti-gravitational
pendants ${\underline A}^{a}$ of the gauge fields 
is introduced via the field strength tensor
\beqn 
{\underline F}^{a}_{{\underline\kappa\underline\nu}} = \nabla_{\underline\kappa} {\underline A}^a_{\underline\nu} 
- \nabla_{\underline\nu} {\underline A}^a_{\underline\kappa}
+ e f^{abc} {\underline A}_{\underline\kappa}^b {\underline A}_{\underline \nu}^c \quad,
\eeqn
where $f^{abc}$ are the structure constants of the group and $e$ is the coupling.
One then constructs the Lagrangian from 
${\underline {\mathcal L}}_A = - \frac{1}{4} {\rm Tr} 
\left( {\underline F}^{\kappa \nu} {\underline F}_{\kappa \nu} \right)$.
Again, a mixture of $F$ with the usual ${\underline F}$ is forbidden by gauge-symmetry. Note, that in all cases the kinetic energy terms are positive.  A particle with negative 
gravitational mass behaves like an ordinary particle except for its
gravitational interaction, encoded in the covariant derivatives.

As one sees by examining the Lagrangian, there is no direct interaction between gravitating and anti-gravitating
particles. 
Nevertheless, both of the particle-species will interact with the gravitational field, which mediates
an interaction between them. However, this coupling is suppressed with the Planck scale.
Therefore, the production of anti-gravitating matter is not observable today because all ingoing particles, 
in whatever scattering process we look at, are the normally gravitating ones that we deal with every day.

\section{Properties of Anti-Gravitating Matter}
   
From the results of the last section, we can now write down the most general form of the
Lagrangian which includes the anti-gravitating matter fields
and is symmetric under exchange of
gravitational with anti-gravitational quantities. It takes the form
\beqn
S &=& \int {\rm d}^{d+4}x~ \sqrt{-g} \left[   G~R 
+  {\mathcal L} + {\underline {\mathcal L}}
\right] 
\label{full} \quad,
\eeqn
where $G=1/m_p^2$, ${\mathcal L} = {\mathcal L}_F +
{\mathcal L}_A$, and ${\underline {\mathcal L}}= {\underline {\mathcal L}}_F +
{\underline {\mathcal L}}_A $.

After a variation of the metric, $\delta g_{\kappa \nu}$, the field equations read
\beqn
R_{\kappa \nu} -\frac{1}{2} R g_{\kappa \nu} = 
\frac{1}{G} \left( T_{\kappa \nu} +  {\underline T}_{\kappa \nu} \right) \quad,
\eeqn 
with the stress-energy tensors ({\sc SET}s) as source terms
\beqn
 T^{\kappa \nu} &=& \frac{2}{\sqrt{-g}}\frac{\delta}{\delta g_{\kappa \nu}}  \left( \sqrt{-g}  {\mathcal L} \right)    
 \quad,\quad
{\underline T}^{\kappa \nu} = \frac{2}{\sqrt{-g}}  \frac{\delta}{\delta g_{\kappa \nu}}  
  \left( \sqrt{-{g}} {\underline {\mathcal L}}  \right) \quad.   
\eeqn  
Using Eq.(\ref{llu}), and 
$g_{\kappa \nu} = \tau^{\kappa}_{\;\;\underline \kappa} \tau^{\nu}_{\;\;\underline \kappa} 
{\underline g}_{\underline \kappa \underline \nu}$,    
and performing the functional derivative, we find 
\beqn
\frac{\delta {\mathcal L}}{\delta g_{\kappa \nu}} = \tau^\kappa_{\;\; {\underline \kappa}} \tau^\nu_{\;\;{\underline \nu}} \frac{\delta {\underline {\mathcal L}}}{\delta {\underline g}{}_{{\underline \kappa} {\underline \nu}}} \quad. \label{dings}
\eeqn
We recall that $\tau_{\kappa {\underline \nu}}$ just transpones quantities and is independent
of the metric tensor. It does not rotate the basis relative to each other, nor 
does it change the volume element or introduce a shear.  
Under a variation of the metric, this quantity remains unaltered $\delta \tau_{\kappa {\underline \nu}} = 0$. 
This, however, is not the case for $\tau_{\kappa}^ {\;\;{\underline \nu}}$, because it transforms non trivially
under coordinate transformations which form a subset of the metric variations.
In particular with Eq. (\ref{cycle}) one obtains
\beqn
\tau^{\kappa {\underline \nu}}~ \tau_{\nu {\underline \kappa}}~
 \delta \left( {\underline g}{}_{{\underline \kappa} {\underline \nu}}~  
  g_{\nu \epsilon} \right) = 0 \quad, \label{minus} 
\eeqn
and, after contraction with $\tau_{\kappa {\underline \epsilon}}~ \tau_{\alpha}^{\;\; {\underline \epsilon}}$,  this 
results in
\beqn
\delta {g}_{\alpha\epsilon} + 
\tau_\alpha^{\;\; {\underline \epsilon}}~ \tau_\epsilon^{\;\;{\underline \kappa}}~  \delta 
{\underline g}{}_{{\underline \kappa} {\underline \epsilon}} = 0 \quad. \label{var}
\eeqn
Inserting Eq. (\ref{var}) in Eq. (\ref{dings}) yields
\beqn
\frac{\delta {\mathcal L}}{\delta g_{\kappa \nu}} = -  \frac{\delta {\underline {\mathcal L}}}{\delta {g}_{\kappa \nu}} \quad.
\eeqn

But the {\sc SET} consists of two terms, the second one arising from the variation of
the volume element.  Using
$
\delta \sqrt{-g} = -\frac{1}{2} g_{\kappa \nu} \sqrt{-g}~ \delta g^{\kappa \nu}  
$,
one finds  the {\sc SET}s for the fields
\beqn
 T^{\kappa \nu} &=&   \frac{\delta {\mathcal L}}{\delta {g}_{\kappa \nu}} - \frac{1}{2} g^{\kappa \nu} {\mathcal L} \quad,\quad
{\underline T}^{\kappa \nu} =  - \tau^\kappa_{\;\; {\underline \kappa}}  \tau^\nu_{\;\; {\underline \nu}} 
\left( \frac{\delta {\underline {\mathcal L}}}{\delta {\underline g}_{{\underline \kappa} {\underline \nu}}} + 
\frac{1}{2} {\underline g}^{{\underline \kappa} {\underline \nu}} 
{\underline {\mathcal L}} \right) \label{sign} \quad.
\eeqn
The interpretation of the so derived gravitational {\sc SET}s is straightforward: Under a perturbation
of the metric, the anti-gravitational fields will undergo a transformation exactly opposite to these of the normal
fields as one expects by construction. The $\tau$-functions convert the indices and the transformation behavior 
from ${\underline{TM}}$ to the usual tangential space.

Most importantly, we see that the {\sc SET} of the anti-gravitating field yields a contribution
to the source of the field equations with a minus sign (and thus justifies the name anti-gravitation). This 
is due to the modified transformation behavior of the field components. However, we also see that the second term, 
arising from the variation of the metric
determinant, does not change sign. For one of the most important cases however, the fermionic matter 
field, the second term vanishes since the Lagrangian is zero when the field fulfills the equation of motion. 

If one considers the Lagrangian of the Dirac-field  one has to formulate the action in form of the tetrad fields.
The above used 
argument then directly transfers to the Dirac field through the properties of the anti-gravitational field under
diffeomorphisms. For the fermions, the {\sc SET} then can be simplified inserting
that the field fulfills the equations of motion $D\hspace*{-.25cm}/\hspace*{.15cm} \Psi = D\hspace*{-.25cm}/\hspace*{.15cm} 
\underline \Psi =0 $. 

The corresponding conservation law of the derived source terms as follows from the Bianchi-identities is as
usual $\nabla^\nu ( T_{\kappa \nu} + {\underline T}_{\kappa \nu}) = 0$.

By variation of the action Eq. (\ref{full}) with respect to the fields, on obtains the equations of
motion in form of the Euler-Lagrange equations. Using the 
the covariant form of Gauss' law one finds as usual
\beqn
\nabla_\nu \left( \frac{\partial {\mathcal L}}{\partial (\nabla_\nu \Psi)}\right) - 
\frac{\partial {\mathcal L}}{\partial \Psi} = 0 \quad.
\eeqn
In complete analogy to the usual case one derives the equations of motion for the anti-gravitational
field
\beqn
{ \nabla}_{\nu} \left( \frac{\partial {\underline {\mathcal L}}}{\partial ({\nabla}_{ \nu} 
{\underline \Psi})}\right) - 
\frac{\partial {\underline {\mathcal L}}}{\partial {\underline \Psi}} = 0 \quad. \label{elag}
\eeqn 
For this derivation one can still use the usual covariant form of Gauss' law since the 
derivatives $\nabla_\nu \underline \Psi$ transform as usual vectors with respect to the index $\nu$.
 
Now it is crucial to note that the kinetic {\sc SET} as defined from the Noether current does {\sl not} have
a change in sign, as no variation of the metric is involved and the gravitational properties of the fields
do not play a role. To clearly distinguish this canonical {\sc SET} from the gravitational
source term, let us denote the canonical {\sc SET} with $\Theta_{\nu \kappa}$, whereas we keep
the above used $T_{\nu \kappa}$  for the gravitational {\sc SET}.

The canonical {\sc SET} for a matter Lagrangian
${\mathcal L}(\Psi, \nabla_\nu \Psi)$ as follows from Noether's theorem is 
\beqn
\Theta^{\nu}_{\;\; \kappa} = \frac{\partial {\mathcal L}}{\partial (\nabla_\nu \Psi) } \nabla
_\kappa \Psi - \delta^{\nu}_{\;\; \kappa} {\mathcal L} \quad,
\eeqn
and is covariantly conserved $\nabla_\nu \Theta^{\nu}_{\;\;\kappa} = 0$. 
Correspondingly one finds the conserved current for the anti-gravitational field
\beqn
{\underline \Theta}^{\nu}_{\;\; \underline \kappa} = 
\tau^{\kappa}_{\;\;\underline \kappa} 
\left( \frac{\partial {\mathcal {\underline L}}}{\partial ({\nabla}_\nu {\underline \Psi}) } 
{\nabla}_{\kappa} {\underline \Psi} - \delta^{\nu}_{\;\;\kappa} 
{\mathcal {\underline L}} \right) \quad,
\eeqn
which is also covariantly conserved $\nabla_\nu \underline \Theta^{\nu}_{\;\;\underline \kappa} = 0$. 

Unfortunately, in general these quantities are neither symmetric, nor are they traceless or gauge-invariant. 
For {\sc GR}, Belinfante's tensor is the more adequate one \cite{enmomt1}. Though the 'correct'
kinetic {\sc SET} is a matter of ongoing discussion \cite{enmomt1,enmomt2,Babak:1999dc}, 
the details will not be important for the following. Instead, let us note that, from the 
Noether current, we get a  total conserved quantity for each space-time direction $\kappa$, leading 
to the conservation equations
\beqn
\nabla^{\nu}  \Theta_{\nu \kappa} + \tau_\kappa^{\;\; \underline \kappa}~ 
 {\nabla}^{\nu} {\underline \Theta}_{\nu \underline\kappa} = 0 \quad. \label{cons2}
\eeqn 

The form of the second term of Eq. (\ref{cons2}) is readily interpreted: when the anti-gravitating 
particle gains kinetic energy 
on a world line, the gravitational particle would loose energy when traveling on the same world line. 
The interaction with the gravitational field is inverted.  
 
One thus can identify the anti-gravitating particle as a particle whose kinetic momentum vector transforms
under general diffeomorphism according to Eq.(\ref{trafo}), whereas the standard particle's kinetic 
momentum transforms according to Eq.(\ref{stdtrafo}). 


It is also instructive to look at the motion of a classical test particle by considering  
the analogue of parallel-transporting the tangent vector. The particle's world line is
denoted ${x}_\nu (\lambda)$, and the anti-gravitating particle's world line is denoted 
${\underline x}_\nu (\lambda)$\footnote{A word of caution is necessary for this notation: the underlined
${\underline x}_\nu$ does only indicate that the curve belongs to the anti-gravitating particle; it is
not related to the curve of the particle ${x}_\nu$.}. 
   
In contrast to the gravitating particle,
the anti-gravitating particle parallel transports not its tangent vector ${\underline t}^\alpha = d{\underline x}^{ \alpha} /d\lambda$, 
but instead the related quantity in ${\underline {TM}}$,  which corresponds to the kinetic momentum, and is
${\underline t}^{\underline \alpha}=  \tau_\alpha^{\;\;\underline \alpha} {\underline t}^{\alpha}$.  On
the particle's world line ${\underline x}_\nu (\lambda)$, it is ${\underline t}^{\underline \alpha}$ which
is covariantly conserved. 
Parallel transporting is then expressed in evaluating the derivative in direction 
of the curve and set it to zero. For the
usual geodesic which parallel transports the tangential vector one has
$t^\nu  
{\nabla}_{\nu}~ {t}^{\alpha} = 0  
$, 
whereas for the anti-gravitating particle one has
\beqn
{\underline t}^{\nu} ~\tau^\alpha_{\;\; \underline\alpha}~ 
{\nabla}_{\nu}~ \tau_\kappa^{\;\; \underline \alpha}~ {\underline t}^\kappa = 0\quad, \label{agpt}
\eeqn
which agrees with the usual equation if and only if the covariant derivative 
on $\tau_\kappa^{\;\; \underline \alpha}$ vanishes. It is important to note that the tangent 
vector ${\underline t}^\alpha$ is not parallel transported along the curve given by the new Eq. (\ref{agpt}).

Using  the covariant derivative  
${\nabla}_{\nu} {\underline t}^{\underline \alpha} = 
{\partial}_{\nu} {\underline t}^{\underline \alpha}
+   
\Gamma^{{\underline \alpha}}_{\;\; {\nu}{\underline \epsilon}}~ 
{\underline t}^{\underline \epsilon}$, 
from 
Eq.(\ref{stdder4}) 
one obtains
\beqn
 \frac{d {\underline x}^{\nu}}{d \lambda} 
\cdot \left( \partial_{\nu} \frac{d {\underline x}^{{\underline \alpha}}}{d \lambda} 
  +  
\Gamma^{\underline \alpha}_{\;\; \nu \underline \epsilon } \frac{d {\underline x}^{\underline \epsilon}}{d \lambda} 
\right) = 0 \quad,
\eeqn
and by rewriting $\partial_{\nu}= (d \lambda /d {\underline x}^{\nu})~ d/d\lambda$ one 
 finds the anti-geodesic equation
\begin{eqnarray} \label{Antigeo}
\frac{d^2 {\underline x}^{{\underline \alpha}}}{d\lambda^2} + \tau^\nu_{\;\;\underline \nu}
\Gamma^{\underline \alpha}_{\;\; \nu \underline \epsilon}
\frac{d {\underline x}^{{\underline \epsilon}}}{d\lambda}
\frac{d {\underline x}^{{\underline \nu}}}{d \lambda} =0 \quad.
\end{eqnarray} 
This equation should be read as an equation for the quantity ${\underline t}^{\underline \alpha}$ rather than an equation
for the curve. To obtain the curve, one proceeds as follows
\begin{itemize}
\item Integrate Eq. (\ref{Antigeo}) once to obtain ${\underline t}^{\underline \alpha}$,
\item Translate this into the geometric tangential vector ${\underline t}^{\underline \alpha}=  \tau_\alpha^{\;\;\underline \alpha} {\underline t}^\alpha$,
\item Integrate a second time with appropriate initial conditions to obtain ${\underline x}^{\alpha}$.
\end{itemize}
Alternatively, one can reformulate Eq. (\ref{agpt}) directly into an equation for the tangential vector
\beqn
{\underline t}^{\nu} ~{\nabla}_{\nu}~  {\underline t}^\alpha + 
{\underline t}^{\nu} {\underline t}^\kappa ~\tau^\alpha_{\;\; \underline\alpha}~  
  {\nabla}_{\nu}~  \tau_\kappa^{\;\; \underline \alpha}
= 0\quad, \label{agpt2}
\eeqn
and insert Eq.(\ref{dtau}).

It will be instructive to also derive this anti-geodesic equation in a second way, 
which makes use of the energy conservation rather than
postulating parallel transport.  
Let us look at an anti-gravitating test particle of nonzero but negative gravitational mass $m$  
moving in a strong gravitational field.  
The particle's energy conservation law  Eq.(\ref{cons2}) in the background field 
reads:
 \begin{eqnarray}
\partial_{\nu} ~{\underline \Theta}^{\nu {\underline \kappa} }
+  
\Gamma^{\underline \kappa}_{\;\; \nu \underline \epsilon} {\underline \Theta}^{ \nu {\underline \epsilon}} 
+     \Gamma^{\nu}_{\;\; \nu \epsilon}
{\underline \Theta}^{\epsilon {\underline \kappa}} = 0 \quad . \label{testT}
\end{eqnarray}

The particle moves on the world line ${y}^{\nu}={ x}^{\nu}(\lambda)$, 
where $\lambda={ x}^0(\lambda)$ is the eigen time. For the
particle of nonzero mass it can be used to parameterize the curve. 
The kinetic {\sc SET} is then  a function of $y$ and can be written as \cite{Weinberg}
\begin{eqnarray} \label{Eimp}
{\underline \Theta}^{\nu \underline \kappa} (y) &=& m \int
\frac{dx^\nu}{d\lambda} 
\frac{dx^{\underline \kappa}}{d\lambda} 
\frac{\delta^4({ y}^{{\epsilon}}-{x}^{{\epsilon}}(\lambda))}
{\sqrt{-g}} d\lambda 
\quad .
\end{eqnarray}           
Taking the partial derivative with respect to $y_\nu$  and rewriting the derivative on the $\delta$-function
yields
\beqn	
\frac{\partial}{\partial y^\nu} {\underline \Theta}^{\nu \underline \kappa}(y)
&=&  -  
m \int  \frac{d\lambda}{\sqrt{-g}} 
 \delta^4({ y}^{{\epsilon}}-{x}^{{\epsilon}}(\lambda))   
  \left[ \frac{d^2 x^{\underline \kappa}}{d\lambda^2} - \Gamma^{\epsilon}_{\;\;\epsilon \nu} 
\frac{dx^\nu}{d\lambda} 
\frac{dx^{\underline \kappa}}{d\lambda} \right]  \label{weiter}
\quad.
\eeqn	
Inserting Eq.(\ref{testT}) yields
\begin{eqnarray} 
0 &=&
 \int \frac{ d\lambda}{\sqrt{-g}} 
 \delta^4({ y}^{{\epsilon}}-{x}^{{\epsilon}}(\lambda))  
\bigg[
 \frac{d^2 x^{\underline \kappa}}{d\lambda^2} - \Gamma^{\epsilon}_{\;\; \epsilon\nu}
\frac{dx^\nu}{d\lambda} 
\frac{dx^{\underline \kappa}}{d\lambda} +  
 \Gamma^{\underline \kappa}_{\;\; \nu \underline \epsilon} 
\frac{dx^\nu}{d\lambda} 
\frac{dx^{\underline \epsilon}}{d\lambda} 
+     \Gamma^{\nu}_{\;\; \nu \epsilon}
\frac{dx^\epsilon}{d\lambda} 
\frac{dx^{\underline \kappa}}{d\lambda}
 \bigg] \nonumber,
\end{eqnarray}	
where the second and the last term in the brackets cancel. Demanding that this be valid on the world 
line of the particle, one again finds Eq.(\ref{Antigeo}). 

On both our ways to derive this equation, we have not used one of the
most common approaches which introduces the particle's world line as the extremal of a variation of the length of
the curve. Here we have instead solely used the consequences from a variation of
the action Eq. (\ref{full}) which includes geodesic motion
as well as anti-geodesic motion.  

It is important to note that the equations of motion Eq.(\ref{Antigeo}) are invariant under general
diffeomorphism, provided that the quantities are transformed appropriately. In case the space-time is
globally flat, one can choose $g=\eta$. It is then also ${\underline g} =\eta$, and $\tau =$ Id, and both sets of
Christoffelsymbols vanish. Since in this case both curves which describe the motion of the particles are identical, 
they will be identical for all choices of coordinate systems in a globally flat background.

\section{Discussion}
\label{discuss}

In the presence 
of strong curvature effects, one expects the interaction between both types
of matter to become important. Since the anti-gravitating contribution to the source-term of the
field equations is negative, the positive energy theorem can be violated. The implications of this
feature for the possibility of singularity avoidance should be investigated further.

Furthermore, the existence of negative gravitational sources can allow gravity to be neutralized, 
which could be used the address the stabilization and compactification problem within the context of 
extra dimensions.

It should also be noted that the anti-gravitating particles do not alter the Hawking-radiation
of black holes. The black hole horizon is a trapped surface only for usual photons, not for
the anti-graviating ones. Indeed, it would be very puzzeling if a particle could be trapped
by a source it is repelled by. Therefore, the anti-gravitating particles will not contribute 
to the Hawking flux. 

\section{Conclusions}
\label{concl}
We have introduced particles into the Standard Model that can cause negative gravitational sources. 
These particles are defined by their transformation behavior under general diffeomorphism. In flat
space they obey the laws of Special Relativity. We thereby have relaxed the equivalence principle.

The number of particles in the Standard Model is doubled in this scenario: each particle has an anti-gravitating partner 
particle which only differs in its opposite reaction to the gravitational field. It is not necessary to
have a negative kinetic energy term. We have shown that the interaction between gravitating and
anti-gravitating matter is mediated solely by gravitation. It is therefore  naturally very weak, explaining
why we have not seen any anti-gravitating matter so far.

\section*{Acknowledgments}
 
I thank Dharam Vir Ahluwalia-Khalilova, 
Marcus Bleicher, Zacharia Chacko, Keith Dienes, Hock Seng Goh, Petr Hajicek, Jim Hartle, 
Stefan Hofmann, Ben Koch, J\"org Ruppert, Stefan Scherer and Lee Smolin for valuable discussions. 
I also thank the Frankfurt Institute of Advanced Studies {\sc (FIAS)} for kind
hospitality. This work was temporarily supported by NSF PHY/0301998, later by the 
Department of Energy under Contract DE-FG02-91ER40618 and the {\sc DFG}.


\begin{thebibliography}{99}


\bibitem{linde1} A.~D.~Linde, Phys.Lett. {\bf B 200}, 272 (1988).

\bibitem{linde2}  
  A.~Linde,
  [arXiv:hep-th/0211048].

  

\bibitem{Moffat:2005ip}
  J.~W.~Moffat,
  [arXiv:hep-th/0507020].


\bibitem{Kaplan:2005rr}
  D.~E.~Kaplan and R.~Sundrum,
  [arXiv:hep-th/0505265].

\bibitem{Bondi}   
H.~Bondi, Rev. Mod. Phys. 29, 423-428 (1957).

\bibitem{Quiros:2004ge}
  I.~Quiros,
  [arXiv:gr-qc/0411064].

\bibitem{Borde:2001fk}
  A.~Borde, L.~H.~Ford and T.~A.~Roman,
  Phys.\ Rev.\ D {\bf 65}, 084002 (2002)
  [arXiv:gr-qc/0109061].


\bibitem{Davies:2002bg}
  P.~C.~W.~Davies and A.~C.~Ottewill,
  Phys.\ Rev.\ D {\bf 65}, 104014 (2002)
  [arXiv:gr-qc/0203003].
 
\bibitem{Ray:2002ts}
  S.~Ray and S.~Bhadra,
  Int.\ J.\ Mod.\ Phys.\ D {\bf 13}, 555 (2004)
  [arXiv:gr-qc/0212120].
  
\bibitem{Rosenberg:2000cv}
  D.~E.~Rosenberg,
  [arXiv:astro-ph/0008166].
 

\bibitem{Torres:1998cu}
  D.~F.~Torres, G.~E.~Romero and L.~A.~Anchordoqui,
  Mod.\ Phys.\ Lett.\ A {\bf 13}, 1575 (1998)
  [arXiv:gr-qc/9805075].
  
  
\bibitem{Zhuravlev:2004vd}
  V.~M.~Zhuravlev, D.~A.~Kornilov and E.~P.~Savelova,
  Gen.\ Rel.\ Grav.\  {\bf 36} (2004) 1719.

\bibitem{Faraoni:2004is}
  V.~Faraoni,
  Phys.\ Rev.\ D {\bf 70}, 081501 (2004)
  [arXiv:gr-qc/0408073].
     
 
 
  
\bibitem{Arkani-Hamed:2003uz}
  N.~Arkani-Hamed, P.~Creminelli, S.~Mukohyama and M.~Zaldarriaga,
  JCAP {\bf 0404}, 001 (2004)
  [arXiv:hep-th/0312100].
\bibitem{Arkani-Hamed:2003uy}
  N.~Arkani-Hamed, H.~C.~Cheng, M.~A.~Luty and S.~Mukohyama,
  JHEP {\bf 0405}, 074 (2004)
  [arXiv:hep-th/0312099].
\bibitem{Arkani-Hamed:2005gu}
  N.~Arkani-Hamed, H.~C.~Cheng, M.~A.~Luty, S.~Mukohyama and T.~Wiseman,
  [arXiv:hep-ph/0507120].
\bibitem{Mann:2005jz}
  R.~B.~Mann and J.~J.~Oh,
  [arXiv:hep-th/0504172].

\bibitem{Krause:2004bu}
  A.~Krause and S.~P.~Ng,
  [arXiv:hep-th/0409241].

 
  
  
\bibitem{enmomt1} R.~E.~G.~Saravi, 
J.~Phys. A 37 (2004) 9573-9586, [arXiv:math-ph/0306020].

 
\bibitem{enmomt2} M.~Forger and H.~R\"omer,
Ann.~Phys.~309 (2004) 306-389, [arXiv:hep-th/0307199].


\bibitem{Babak:1999dc}
  S.~V.~Babak and L.~P.~Grishchuk,
  Phys.\ Rev.\ D {\bf 61}, 024038 (2000)
  [arXiv:gr-qc/9907027].
  

\bibitem{Weinberg} S.~Weinberg, '{\sl Gravitation and Cosmology }':  Wiley (July, 1972) .
 
 
 
\end{thebibliography}
\end{document}